\documentclass[aps,prl,twocolumn,groupedaddress,showpacs]{revtex4}

\usepackage{graphicx}
\usepackage{dcolumn}
\usepackage{bm}

\begin{document}

\title{Orbital Ordering, New Phases, and Stripe Formation
in Doped Layered Nickelates}

\author{Takashi Hotta$^1$ and Elbio Dagotto$^2$}

\affiliation{$^1$Advanced Science Research Center,
Japan Atomic Energy Research Institute,
Tokai, Ibaraki 319-1195, Japan \\
$^2$National High Magnetic Field Laboratory,
Florida State University, Tallahassee, Florida 32306}

\date{\today}

\begin{abstract}
Ground-state properties of layered nickelates are
investigated based on the orbital-degenerate Hubbard model
coupled with lattice distortions, by using numerical techniques.
The N\'eel state composed of spin $S$=1 ions is confirmed
in the undoped limit $x$=0.
At $x$=1/2, novel antiferromagnetic states, called
CE- and E-type phases, are found by increasing the Hund's coupling.
($3x^2$$-$$r^2$/$3y^2$$-$$r^2$)-type orbital ordering is
predicted to occur in a checkerboard-type charge-ordered state.
At $x$=1/3, $both$ Coulombic and phononic interactions are found
to be important, since the former stabilizes the spin stripe,
while the latter leads to the striped charge-order.
\end{abstract}



\maketitle


The existence and origin of ``striped'' structures continues attracting
considerable attention in the research field of
transition-metal oxides \cite{Tranquada}.
In a system with dominant electron-electron repulsion,
the Wigner-crystal state should be stabilized,
but in real materials more complicated non-uniform charge
structures have been found.
In Nd-based lightly-doped cuprates,
neutron scattering experiments revealed incommensurate
spin structures \cite{neutron} where antiferromagnetic (AFM)
spin stripes are periodically separated by domain walls of holes.
In $\rm La_{2-{\it x}}Sr_{\it x}CuO_4$, dynamical stripes are believed to exist
along vertical or horizontal directions (Cu-O bond direction)
\cite{Matsuda}. In nickelates, the charge-ordered stripes  are along
the diagonal direction \cite{Ni-stripe}.
In manganites, evidence for striped charge-ordering also along the diagonal
direction has been reported in the AFM phase for $x$$>$1/2 \cite{Mori},
while short-range diagonal stripe correlations have been found
in the ferromagnetic (FM) phase at $x$$<$1/2 \cite{Dai}.

In general, stripes can be classified into metallic or insulating.
In $\rm La_{2-{\it x}}Sr_{\it x}CuO_4$, the dynamical stripes
exhibit metallic properties, but they are easily pinned
by lattice effects and impurities.
In $\rm La_{1.6-{\it x}}Nd_{0.4}Sr_{\it x}CuO_4$, stripes along
the bond-direction are pinned by lattice distortions
\cite{Tranquada}, but they are still metallic.
Intuitively, vertical or horizontal
stripes could be associated with the formation of ``rivers of holes'',
to prevent individual charges from fighting against
the AFM background \cite{pi-shift}.
Such stripes should be metallic, even if they are pinned,
since they are induced by the optimization of hole motion
between nearest-neighbor Cu-sites via oxygens.

However, in the diagonal stripes observed in manganites and nickelates,
charges are basically localized,
indicating that such insulating stripes are $not$ determined
just by the optimization of the hole motion.
In the FM state of manganites, the hole movement is already
optimal and, naively, charges should not form stripes.
Obviously, an additional effective local potential must be
acting to confine electrons into stripes.
If such a potential originates in lattice distortions, it
is expected to occur along the bond direction
to avoid energy loss due to the conflict
between neighboring lattice distortions sharing the same oxygens.
Then, static stripes stabilized by lattice distortions tends to
occur along the $diagonal$ direction,
as shown in the stripes of the FM-phase of manganites,
stabilized by Jahn-Teller (JT)
distortions \cite{Hotta1}.

In simple terms, vertical or horizontal stripes in cuprates
can be understood by the competition between Coulomb interaction
and hole motion, while diagonal stripes are better explained as a
consequence of a robust electron-lattice coupling.
However, a difficulty has been found
for theoretical studies of stripe formation in doped nickelates,
since both Coulomb interaction and
electron-lattice coupling appear to be important.
Since the Ni$^{2+}$ ion has two electrons in the $e_{\rm g}$ orbitals,
on-site Coulomb interactions certainly play a crucial role to form
spins $S$=1.
When holes are doped, one electron is removed and another remains
in the $e_{\rm g}$ orbitals, indicating that the hole-doped site
should become JT active. Then,
in hole-doped nickelates $both$ Coulombic and phononic interactions
could be of relevance, a fact not considered in previous theoretical
investigations.

In this Letter, charge ordering
in doped nickelates is investigated based on
the orbital-degenerate Hubbard model coupled
to lattice distortions, using numerical techniques.
After confirming the N\'eel state composed of $S$=1 spins
at $x$=0, both cases $x$=1/2 and 1/3 will be analyzed.
At $x$=1/2, novel AFM phases called CE- and E-type have been
unveiled, which are consistent with experimental results.
Further including the JT-type cooperative distortion,
($3x^2$$-$$r^2$/$3y^2$$-$$r^2$)-type orbital ordering is predicted.
For $x$=1/3, an incommensurate spin structure is induced
by the Coulombic model, including the level splitting between $e_{\rm g}$
orbitals, but the charge stripe does not appear.
To reproduce simultaneously spin and charge stripes,
it is important to include the strong coupling of $e_{\rm g}$
electrons to lattice distortions, originating in the in-plane
oxygen motions.


The model for nickelates includes three important ingredients:
The kinetic motion of $e_{\rm g}$ electrons,
Coulomb interactions among $e_{\rm g}$ electrons,
and electron-lattice couplings between $e_{\rm g}$ electrons
and distortions of the NiO$_6$ octahedra.
Note that the electron-lattice term is divided into 
couplings for the apical and in-plane oxygen motions.
In layered nickelates, all NiO$_6$ octahedra are significantly
elongated along the $c$-axis, splitting the
$e_{\rm g}$ orbitals.
This splitting from apical oxygens
should be included explicitly from the start and, then,
the in-plane motion should be studied.
The Hamiltonian $H$ is given by
\begin{eqnarray}
  H &=& -\sum_{{\bf ia}
  \gamma \gamma'\sigma}t^{\bf a}_{\gamma \gamma'}
  d_{{\bf i} \gamma \sigma}^{\dag}d_{{\bf i+a} \gamma' \sigma}
  +\Delta \sum_{\bf i}(n_{{\bf i}{\rm a}}-n_{{\bf i}{\rm b}})/2 \nonumber \\
  &+& U \sum_{{\bf i},\gamma}
  n_{{\bf i}\gamma\downarrow}n_{{\bf i}\gamma\uparrow}
  + J \sum_{{\bf i},\sigma,\sigma'}
  d_{{\bf i}{\rm a}\sigma}^{\dag} d_{{\bf i}{\rm b}\sigma'}^{\dag}
  d_{{\bf i}{\rm a}\sigma'} d_{{\bf i}{\rm b}\sigma} \nonumber \\
  &+& U' \sum_{\bf i}n_{{\bf i}{\rm a}} n_{{\bf i}{\rm b}}
  + J' \sum_{{\bf i},\gamma \ne \gamma'}
  d_{{\bf i}\gamma\uparrow}^{\dag} d_{{\bf i}\gamma\downarrow}^{\dag}
  d_{{\bf i}\gamma'\downarrow} d_{{\bf i}\gamma'\uparrow},
\end{eqnarray}
where $d_{{\bf i}{\rm a}\sigma}$ ($d_{{\bf i}{\rm b}\sigma}$) is the
annihilation operator for an $e_{\rm g}$-electron with spin $\sigma$
in the $d_{x^2-y^2}$ ($d_{3z^2-r^2}$) orbital at site ${\bf i}$,
$n_{{\bf i} \gamma\sigma}$=
$d_{{\bf i} \gamma \sigma}^{\dag}d_{{\bf i} \gamma \sigma}$,
$n_{{\bf i}\gamma}$=$\sum_{\sigma}n_{{\bf i} \gamma\sigma}$,
${\bf a}$ is the vector connecting nearest-neighbor sites,
and $t^{\bf a}_{\gamma \gamma'}$ is the nearest-neighbor hopping
amplitude between $\gamma$- and $\gamma'$-orbitals along the
${\bf a}$-direction, given by
$t_{\rm aa}^{\bf x}$=$-\sqrt{3}t_{\rm ab}^{\bf x}$=
$-\sqrt{3}t_{\rm ba}^{\bf x}$=$3t_{\rm bb}^{\bf x}$=$3t/4$
for the $x$-direction and
$t_{\rm aa}^{\bf y}$=$\sqrt{3}t_{\rm ab}^{\bf y}$=
$\sqrt{3}t_{\rm ba}^{\bf y}$=$3t_{\rm bb}^{\bf y}$=$3t/4$
for the $y$-direction \cite{Dagotto}.
Hereafter, $t$ is the energy unit.
In the second term, $\Delta$($>$0) is the level splitting
between a- and b-orbitals.
In the Coulomb interaction terms,
$U$ ($U'$) is the intra-orbital (inter-orbital) Coulomb interaction,
$J$ is the inter-orbital exchange interaction, and
$J'$ is the pair-hopping amplitude between different orbitals.
Due to the relations $J$=$J'$ and $U$=$U'$+$J$+$J'$,
the independent parameters are $U'$ and $J$, with $U'$$>$$J$
\cite{Dagotto}. The calculations are all carried out using
standard exact-diagonalization techniques.
It is crucial to use this kind of $unbiased$ methods for
this first study that includes both Coulomb and lattice effects,
even if the technique restricts us to small $N$-site clusters.

\begin{figure}[t]
\includegraphics[width=0.9\linewidth]{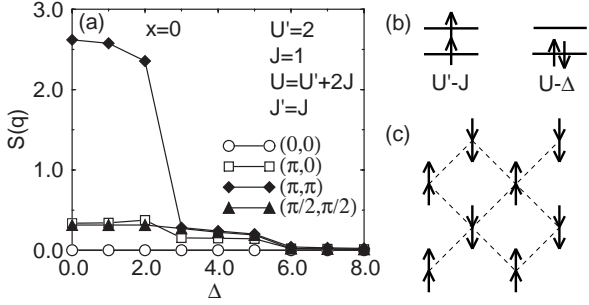}
\caption{(a) Spin correlation $S(\bm{q})$ vs. $\Delta$ for $x$=0.
(b) Two kinds of local $e_{\rm g}$-electron arrangements for $x$=0.
(c) AFM spin pattern theoretically determined for $\Delta$$\alt$3.}
\end{figure}


First, consider the undoped case.
The calculation is done for an 8-site tilted cluster, equivalent
in complexity to a 16-site lattice for the single-band Hubbard model.
Since at all sites the two orbitals are occupied
due to the Hund's rule coupling, the JT distortions are not active
and it is possible to grasp the essential ground-state properties using
$H$. In Fig.~1(a), the Fourier transform of spin correlations
is shown vs. $\Delta$, where
$S(\bm{q})$=$(1/N)$$\sum_{\bf i,j}e^{i\bm{q}\cdot({\bf i}-{\bf j})}
\langle S^{z}_{{\bf i}} S^{z}_{{\bf j}} \rangle$, with
$S^{z}_{{\bf i}}$=
$\sum_{\gamma}(d_{{\bf i} \gamma\uparrow}^{\dag}d_{{\bf i}\gamma\uparrow}$
$-$$d_{{\bf i} \gamma\downarrow}^{\dag}d_{{\bf i}\gamma\downarrow})/2$.
As expected, a robust $(\pi, \pi)$ peak can be observed for
$\Delta$$\alt$3, suggesting that the AFM phase is stabilized by
super-exchange interactions.
The rapid decrease of $S(\pi, \pi)$ for $\Delta$$\agt$3 is understood by
comparing the energies for local triplet and singlet states,
as shown in Fig.~1(b).
The ground-state properties change at $U'$$-$$J$=$U$$-$$\Delta$,
leading to $\Delta$=$3J$ for the transition.
The spin structure at $x$=0 is schematically shown in Fig.~1(c).


Let us turn our attention to the case $x$=1/2.
The 8-site tilted lattice is again used for the analysis,
and the phase diagram Fig.~2(a) is obtained 
for $\Delta$=0.5 \cite{experimental}.
Increasing $J$, an interesting transformation from AFM
to FM phases is found. This is natural, since at large $J$
the system has a formal similarity with manganite models,
where kinetic-energy gains lead to ferromagnetism, while at 
small $J$ the magnetic energy dominates.
However, between the G-type AFM for $J$$\approx$0
and FM phase for $J$$\approx$$U'$, unexpected states appear which are
mixtures of FM and AFM phases, due to the competition between
kinetic and magnetic energies.
Typical spin correlations $S(\bm{q})$ are shown in Fig.~2(b).
Note that peaks at $\bm{q}$=$(\pi, 0)$ and $(\pi/2,\pi/2)$
indicate ``C'' and ``E'' type spin-structures, respectively
(the notation is borrowed from the Mn-oxide context \cite{Dagotto}).
Double peaks at $\bm{q}$=$(\pi, 0)$ and $(\pi/2,\pi/2)$
denote the CE-type structure, frequently observed in half-doped
manganites \cite{murakami}.
In half-doped nickelates, the CE-phase is expressed as
a mixture of type (I) and (II) in Fig.~2(c),
depending on the positions of the $S$=1 and $S$=1/2 sites,
although the ``zigzag'' FM chain structure is common for both types.
The E-type phase is also depicted in Fig.~2(c).
Note that the charge correlation always exhibits a peak
at $\bm{q}$=$(\pi, \pi)$ (not shown here),
indicating the checkerboard-type charge ordering.

In experimental results, a peak at $(\pi/2,\pi/2)$ in $S(\bm{q})$
has been reported \cite{Ni-stripe}, suggesting an AFM pair of
$S$=1 spins $across$ the singly-occupied sites with holes.
Moreover, the checkerboard-type charge ordering
has been experimentally observed \cite{Ni-stripe}.
Thus, the spin-charge patterns of CE(II)- and E-type
are consistent with the experimental results.
Our phase diagram has a robust region with
a peak at $(\pi/2,\pi/2)$, both for CE- and E-type phases,
although the CE-phase exhibits an extra peak at $(\pi, 0)$.
Whether the E- or CE-phases are present in nickelates can be studied 
experimentally in the future by searching for this $(\pi, 0)$ peak.
Note that if diffuse scattering experiments detect the AF correlation
$along$ the hole stripe, as has been found at $x$=1/3 \cite{Boothroyd},
the CE(II)-type may be the only possibility.
Summarizing, the spin-charge structure in $x$=1/2 experiments can be
understood within the Hamiltonian $H$ by assuming a relatively large $J$.


Consider now the effect of in-plane oxygen
motion (apical oxygen motions have already been
included as an $e_{\rm g}$-level splitting).
Assuming that oxygens move along the Ni-O bond direction,
the extra electron-phonon coupling term is written as
\begin{eqnarray}
  H_{\rm eph} &=&
  g \sum_{\bf i}[-Q_{1{\bf i}}(n_{{\bf i}{\rm a}}+n_{{\bf i}{\rm b}})
  +Q_{2{\bf i}} \tau_{x{\bf i}}+Q_{3{\bf i}} \tau_{z{\bf i}}] \nonumber \\
  &+& (k/2) \sum_{\bf i} (\beta Q_{1{\bf i}}^2
  +Q_{2{\bf i}}^2+Q_{3{\bf i}}^2),
\end{eqnarray}
where $g$ is the electron-lattice coupling constant,
$Q_{1{\bf i}}$ is the breathing-mode distortion,
$Q_{2{\bf i}}$ and $Q_{3{\bf i}}$ are, respectively, the
$(x^2$$-$$y^2)$- and $(3z^2$$-$$r^2)$-type JT distortions,
$\tau_{{\rm x}{\bf i}}=
\sum_{\sigma}(d_{{\bf i}{\rm a}\sigma}^{\dag}d_{{\bf i}{\rm b}\sigma}
+d_{{\bf i}{\rm b}\sigma}^{\dag}d_{{\bf i}{\rm a}\sigma})$, and
$\tau_{{\rm z}{\bf i}}=
\sum_{\sigma}(d_{{\bf i} {\rm a}\sigma}^{\dag}d_{{\bf i}{\rm a}\sigma}
-d_{{\bf i}{\rm b}\sigma}^{\dag}d_{{\bf i}{\rm b}\sigma})$.
The second term is the quadratic potential for adiabatic distortions,
where $k$ is the spring constant for the JT-mode and $\beta$ is
the spring-constant ratio for breathing- and JT-modes. From our
experience in manganites, this ratio is here fixed to
$\beta$=2 \cite{Hotta2}.

\begin{figure}[t]
\includegraphics[width=0.95\linewidth]{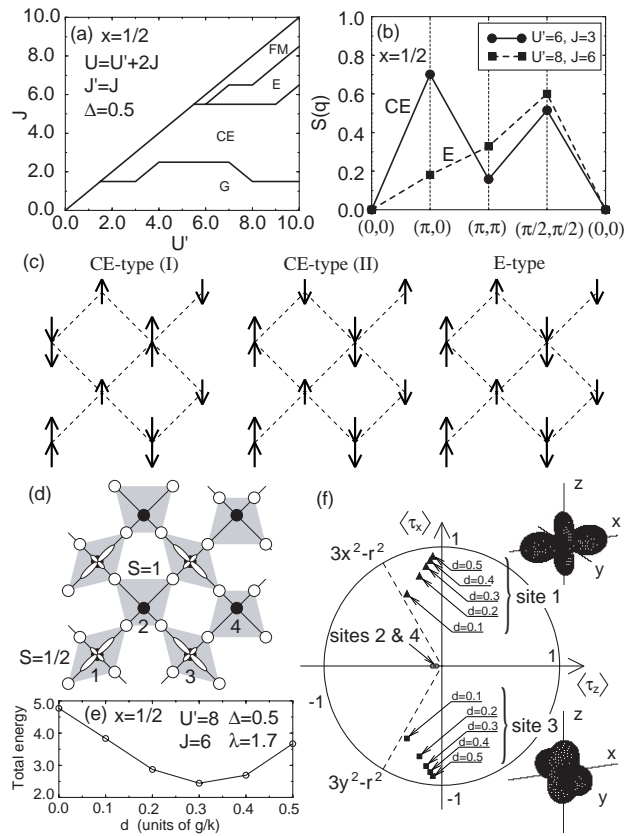}
\caption{(a) Ground-state phase diagram at $x$=1/2.
(b) $S(\bm{q})$ for the CE- and E-type phases, at the couplings indicated.
(c) Spin and charge patterns for the CE- and E-type phases.
These are schematic views, since local charge-densities in practice
are not exactly 1 and 2.
(d) Numerically obtained cooperative distortion pattern for an
8-site lattice at $x$=1/2.
Black and open circles indicate Ni and O ions, respectively.
Open symbols indicate $e_{\rm g}$ orbitals in the optimized state.
(e) Total ground-state energy vs. $d$ for $x$=1/2.
(f) Orbital densities $\langle \tau_{z{\bf i}} \rangle$
and $\langle \tau_{x{\bf i}} \rangle$ for sites 1--4.
See (d) for the site labels.
Optimized orbitals at $d$=0.3 for sites 1 and 3 are also shown.}
\end{figure}

Since all oxygens are shared by adjacent NiO$_6$ octahedra,
the distortions are $not$ $independent$.
To consider such cooperative effect, in principle,
the O-ion displacements should be optimized.
However, in practice it is not
feasible to perform both the Lanczos diagonalization and
the optimization of all oxygen positions for 6- and 8-site clusters.
In the actual calculations, $Q_{1{\bf i}}$, $Q_{2{\bf i}}$, and
$Q_{3{\bf i}}$ are expressed by a single parameter $d$, for the shift
of the O-ion coordinate. Note that the unit of $d$ is $g/k$, typically
0.1$\sim$0.3\AA.
Then, the total energy is evaluated as a function of $d$
to find the minimum energy state.
Repeating these calculations for several distortion patterns,
it is possible to deduce the optimal state.

After several trials, the optimal distortion
at $x$=1/2 is shown in Fig.~2(d).
The diagonalization has been performed at several values of
$d$ on the 8-site distorted lattice and the minimum in the total
energy is found at $d$=0.3 (Fig.~2(e)). Here, the dimensionless
coupling constant $\lambda$ is defined as $\lambda$=$g$/$\sqrt{kt}$.
As mentioned above, even without $H_{\rm eph}$,
the checkerboard-type charge ordering has been obtained,
but the peak at $\bm{q}$=$(\pi, \pi)$ significantly grows
due to the effect of lattice distortions.
Note that the distortion pattern in Fig.~2(d) is essentially
the same as that for half-doped manganites.
This is quite natural, since JT active and inactive ions exist
bipartitely also for half-doped nickelates.
Then, due to this JT-type distortion
{\it orbital ordering for half-doped nickelates} is predicted,
as schematically shown in Fig.~2(d).
The shapes of orbitals are determined from the orbital densities,
$\langle \tau_{z{\bf i}} \rangle$ and
$\langle \tau_{x{\bf i}} \rangle$ (Fig.~2(f)).
The well-known alternate pattern of $3x^2$$-$$r^2$ and
$3y^2$$-$$r^2$ orbitals
in half-doped manganites is denoted by dashed lines.
Increasing $d$, the shape of orbitals
deviates from $3x^2$$-$$r^2$ and $3y^2$$-$$r^2$,
but it is still characterized
by the orbitals elongating along the $x$- and $y$-directions
(see insets of Fig.~2(f)).
It would be very interesting to search for orbital ordering in
half-doped nickelates, using the resonant X-ray scattering technique.


Now let us move to the case $x$=1/3.
If the actual expected stripe structure at $x$=1/3 is faithfully
considered \cite{Ni-stripe},
it is necessary to analyze, at least, a 6$\times$6 cluster.
However, such a large-size cluster with orbital degeneracy
cannot be treated exactly due to the exponential growth of the
Hilbert space with cluster size.
Then, a covering of the two-dimensional (2D) lattice
using zigzag 6-sites clusters is considered
(Fig.~3(a)) by assuming
a periodic structure along the diagonal direction.
The phase diagram obtained by analyzing the zigzag 6-site cluster
for $H$ is in Fig.~3(b).
Typical spin and charge correlations are in Figs.~3(c) and (d), where
$C(\bm{q})$=$(1/N)$$\sum_{\bf i,j}e^{i\bm{q}\cdot({\bf i}-{\bf j})}$
$\langle (n_{\bf i}-\langle n \rangle)$$\cdot$$(n_{\bf j}-
\langle n \rangle) \rangle$,
with $n_{\bf i}$=$\sum_{\gamma}n_{{\bf i}\gamma}$.

Since the momentum $q$ is defined along the zigzag direction,
the phase labelled by $q$=$2\pi/3$ in Fig.~3(b) denotes
an $incommensurate$ $AFM$ $phase$ with the proper spin stripe structure.
The phase labelled by $q$=$\pi/3$ indicates
a spin spiral state, which will eventually turn to
the FM phase in the thermodynamic limit.
Thus, the spin stripe phase appears between the commensurate
AFM and FM-like phases, similar to the case of $x$=1/2.
However, as seen in Fig.~3(d), $C(q)$ in the spin stripe phase
does $not$ show the striped charge structure ($q$=$2\pi/3$).
Rather, bipartite charge ordering characterized by
a peak at $q$=$\pi$ still remains.
Namely, the Hamiltonian $H$ can explain the spin stripe,
but does $not$ reproduce the striped charge ordering at $x$=1/3,
indicating the importance of $H_{\rm eph}$.

\begin{figure}[t]
\includegraphics[width=0.95\linewidth]{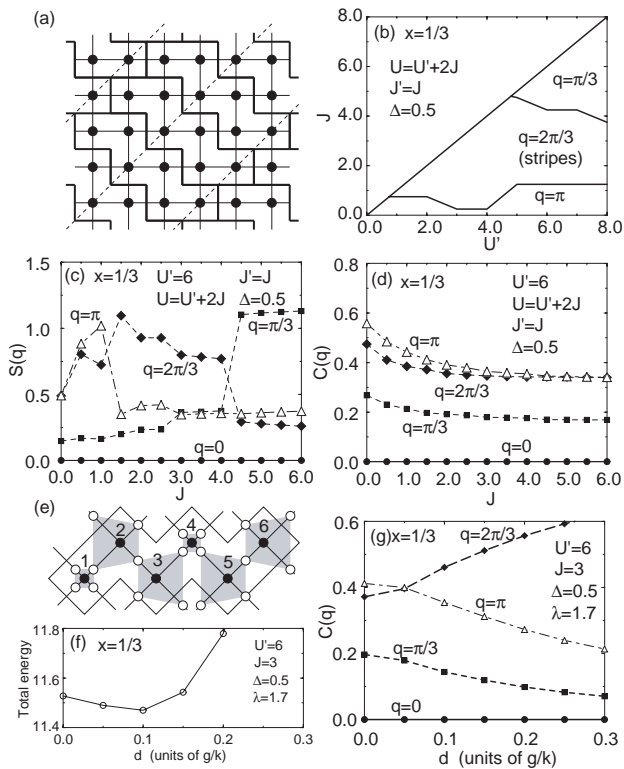}
\caption{(a) Zigzag 6-sites cluster covering the 2D lattice.
Black circles denote Ni ions, and dashed lines indicate hole positions.
(b) Phase diagram at $x$=1/3. Each phase is characterized by
the momentum that shows a peak in $S(q)$.
(c) $S(q)$ and (d) $C(q)$ vs. $J$ for $U'$=6 and $\Delta$=0.5.
(e) Cooperative distortion pattern for the zigzag 6-sites cluster
at $x$=1/3.
(f) Total ground-state energy and (g) $C(q)$ vs. $d$ for $x$=1/3.}
\end{figure}

Consider now the effect of $H_{\rm eph}$ for $x$=1/3.
After evaluating total ground-state energies for several kinds of
distortions, the pattern in Fig.~3(e) has been found to provide
the optimal state at $x$=1/3.
This type of distortion induces a spatial modulation of
the level splitting as
$-\delta_1/2$=$\delta_2$=$\delta_3$=$-\delta_4$/2=$\delta_5$=$\delta_6$
\cite{Tranquada2},
where $\delta_{\bf i}$ is the level splitting caused by the
in-plane oxygen motions, and the site numbers are in Fig.~3(e).
The minimum energy is found at $d$=0.1 (Fig.~3(f)).
The modulation of level splitting stabilizes the striped charge
ordering characterized by a $q$=$2\pi/3$ peak in $C(q)$ (Fig.~3(g)).

Note that ($3x^2$$-$$r^2$/$3y^2$$-$$r^2$)-type orbital ordering
does $not$ occur in Fig.~3(e). Phenomenologically, such orbital
ordering tends to appear in a hole pair separated by one site,
the unit of the ``bi-stripe'' of manganites \cite{Mori}. However,
such a bi-stripe-type ordering contradicts the $x$=1/3 striped 
charge-ordering, and the bi-stripe-type solution was found to be
unstable in these calculations.
One may consider other distortion patterns which satisfy
both ($3x^2$$-$$r^2$/$3y^2$$-$$r^2$)-type orbital and
striped charge-ordering, but in such distortions
no energy minimum was obtained for $d$$>$0. After several trials,
Fig.~3(e) has provided the most optimal state.


Summarizing, possible spin, charge, $and$ orbital structures of
layered nickelates have been discussed
based on the $e_{\rm g}$-orbital degenerate
Hubbard model coupled with lattice distortions.
To understand the nickelate stripes,
{\it both Hund's rule interaction and electron-lattice coupling
appear essentially important}.
At $x$=1/2, ($3x^2$$-$$r^2$/$3y^2$$-$$r^2$)-type
orbital ordering similar to that in half-doped manganites
is predicted. Even FM phases could be stabilized by
chemically altering the carrier's bandwidth.
For $x$=1/3, a spatial modulation in level splitting
plays an important role for stripe formation.


The authors thank M. Matsuda, J. Tranquada, and H. Yoshizawa
for discussions.
T. H. is supported by Priority-Areas Grant from the Ministry of
Education, Culture, Sports, Science, and Technology of Japan.
E. D. is supported by the NSF grants DMR-0122523 and 0312333.



\begin{thebibliography}{99}
\bibitem{experimental}
Since $\Delta$ of nickelates is half of that
of cuprates, from the lattice constants for $\rm CuO_6$ and
$\rm NiO_6$ octahedra, it is reasonable to select $\Delta$/$t$=0.5.
\end{thebibliography}
\end{document}